\documentclass{emulateapj}
\usepackage{times}
\usepackage{graphicx}

\slugcomment{To Appear in ApJL}

\shorttitle{PSR~J0636+5128}
\shortauthors{Draghis \& Romani}

\begin{document}

\title{PSR J0636+5128: A Heated Companion in a Tight Orbit }

\author{Paul Draghis \& Roger W. Romani\altaffilmark{1}}
\altaffiltext{1}{Department of Physics, Stanford University, Stanford, CA 94305-4060,
 USA; rwr@astro.stanford.edu}

\begin{abstract}

	We present an analysis of archival Gemini $g^\prime,~ r^\prime, K$ 
and Keck $H, ~ K_s$ imaging
of this nearby short period binary ($P_B=95.8$\,min) 2.87\,ms pulsar. The heated
companion is clearly detected. Direct pulsar heating provides an acceptable model
at the revised $>700$\,pc parallax distance. The relatively shallow light curve modulation
prefers an inclination $i<40^\circ$; this high latitude view provides
a likely explanation for the lack of radio signatures of wind dispersion or eclipse.
It also explains the low minimum companion mass and, possibly, the faintness of the source 
in X- and $\gamma$-rays.
\end{abstract}


\section{Introduction}

	PSR~J0636+5128 (hereafter J0636) is a $P=2.87$\,ms millisecond pulsar (MSP) binary 
discovered in the Green Bank Northern Celestial Cap Pulsar Survey \citep{stoet14}. 
It was originally designated PSR ~J0636+5129, but improved timing astrometry requires
an official name change. It has the third shortest orbital 
period ($P_B=5750$\,s) of any confirmed pulsar binary, and a small mass function 
indicating a minimum companion mass $m_{c,min}=0.0070M_\odot$. This is likely a
member of the `black widow' class of evaporating pulsars, although no
radio eclipses or orbital modulation of dispersion measure (DM) are seen. 
With $DM=11.1 {\rm cm^{-3}pc}$ one infers a distance of
$\sim 500$\,pc (NE2001, Cordes \& Lazio 2002). A recent timing parallax indicates
$d>700\,$pc \citet{arzet18}, so the 
\citet{stoet14} parallax estimate of $d = 203_{-21}^{+27}$\,pc was evidently in error. 
The proper motion is small and the
Shklovskii-corrected spindown power is ${\dot E} =5.6\times 10^{33} I_{45} {\rm erg\, s^{-1}}$,
for a neutron star moment of inertia of $10^{45} I_{45} {\rm g\,cm^2}$. This is
a relatively low spindown power for a companion-evaporating pulsar. The binary is
detected in the X-rays by Chandra, but must be relatively faint in the
$\gamma$-rays as it is not in the 8-y {\it Fermi} LAT catalog.

	We found observations of the field of the pulsar in the Keck 
and Gemini archives. The companion is clearly detected at the radio-determined
position, is brightest at the expected maximum phase and substantially variable (Figure 1).  
We used these data to measure and fit the companion light curve.

\section{Archival data -- Optical GMOS Images}
	
	The Gemini Science archive contained ten 420\,s $g^\prime$ exposures and nine
420\,s $r^\prime$ images, taken on December 21, 2014 under program GN-2014B-Q-81
(PI Stovall). Images in the two bands had FWHM median 0.91$^{\prime\prime}$ and 0.89$^{\prime\prime}$,
respectively.  We downloaded these data along with associated calibration frames and standard fields 
from the Gemini archive and subject them to standard Gemini IRAF calibrations. All frames were 
registered to a common position and
we performed Gaussian-weighted aperture photometry at the maximum light pulsar position.
After some experimentation to optimize the S/N, we adopted a
Gaussian weight $\sigma$ of $1.4\times$ the individual frame's FWHM and an extraction
aperture $1.7\times$ this FWHM; we used a local annular background. We calibrated 
by using matched weighted extractions of field stars, whose $g^\prime$ and $r^\prime$ 
magnitudes were determined from the SDSS catalog.  
The pulsar was detected in all frames with reasasonable statistical significance. However
a faint extended source, likely a background galaxy, lies $\sim 1.5^{\prime\prime}$ to the
SW (Figure 1, lower left). In the $g^\prime$ frame at minimum, which suffered relatively 
poor seeing, this may provide a systematic contamination, so our measurement might 
best be considered an upper limit (Figure 2).
We estimate the zero point errors as $\le 0.03$mag, insignificant
except for the very highest S/N points.

	After barycentering the time of the frame midpoints, we plotted the measured magnitudes
folded on the orbital ephemeris of \citet{stoet14} (Figure 2). The source is brightest at pulsar
inferior conjunction ($\phi_B=0.75$) when we are best viewing the heated face of the companion.
The orbital coverage is incomplete, and there appear to be fluctuations about the over-all
quasi-sinusoidal modulation. 

\begin{figure}[h!!!!]
\vskip 5.5truecm
\includegraphics{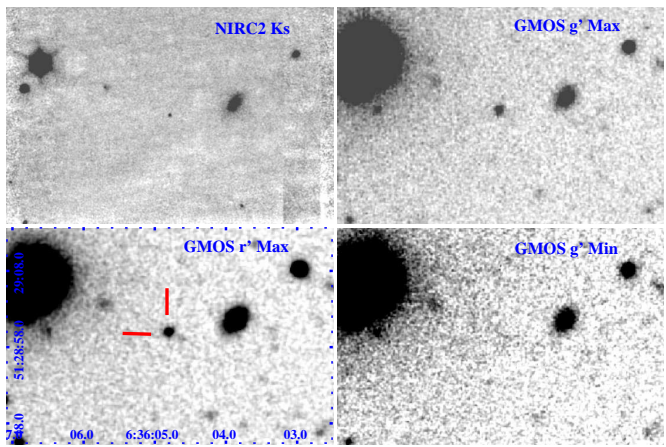}
\begin{center}
\caption{\label{MaxMin} 
Upper Left: NIRC2 $K_s$ image of PSR J0636+5128 (median of frames). Lower Left: GMOS $r$ image at maximum.
Upper Right: GMOS $g$ image at maximum. Lower Right: GMOS $g$ image at minimum.
}
\end{center}
\vskip -0.8truecm
\end{figure}

\section{Archival data -- Near-IR Gemini NIRI and Keck NIRC2 }

	The same Gemini program includes 60\,s $K$ integrations, taken with NIRI+AO+LGS on 
November 05 (21 exposures) and December 08 2014 (24 usable exposures).
The typical (AO-assisted) FWHM were 0.19$^{\prime\prime}$ and $0.14^{\prime\prime}$ on the two nights.
Again optimal extraction provided estimated photometry for the two nights. 
The $K$ magnitudes were calibrated against the $K_s=13.88$ star
2MASS J06360673+5129070. In a few frames this star was saturated or too near the image edge and so
we bootstrapped the calibration via objects common to the unsaturated frames.

	The Keck archive contained NIRC2+AO frames of the field (program C222N2L, Kulkarni, PI)
with 27$\times$60\,s in $H$ and 28$\times$60\,s in $K^\prime$ taken on March 1, 2013. 
The frame FWHM was $0.18-0.21^{\prime\prime}$ in $H$ and $0.14-0.19^{\prime\prime}$ in $K^\prime$.
For these data relative unweighted aperture photometry proved most stable. Again we calibrated against
2MASS J06360673+5129070, correcting for the slow transparency variations through the observation.

	The near-IR fluxes still showed occasional outliers, even after this flux calibration.
Accordingly we chose to adopt the median of each set of three images in the IR pointing dithers.
In Figure 2 The near-IR light curves show the individual points in the first cycle and the median points
in the second period.  Together the Gemini and Keck $K$ magnitudes cover nearly a full orbit. 
Where these overlap, the Keck magnitudes appear fainter, but with substantial scatter.  Given the evident
variation about the sinusoidal light curve, we cannot differentiate between photometric errors
or companion brightening. The overall $K$ light curve, however has a shallow minimum at phase 
$\phi=0.25$.

	With $\Delta m < 2$\,mag the optical light curves are relatively shallow for a 
classical black widow. However, the the deeper modulation in the $g^\prime$ band gives the
signature of companion heating. The light curves show evidence of non-sinusoidal
variability in all four bands. Some short period companion-evaporating pulsars show stochastic
variability, likely due magnetic flares on the companions \citep{ret15}. So until
we have many orbits of J0636 photometry we cannot establish a full quiescence 
light curve. 

\begin{figure}[t!!]
\vskip 8.5truecm
\includegraphics{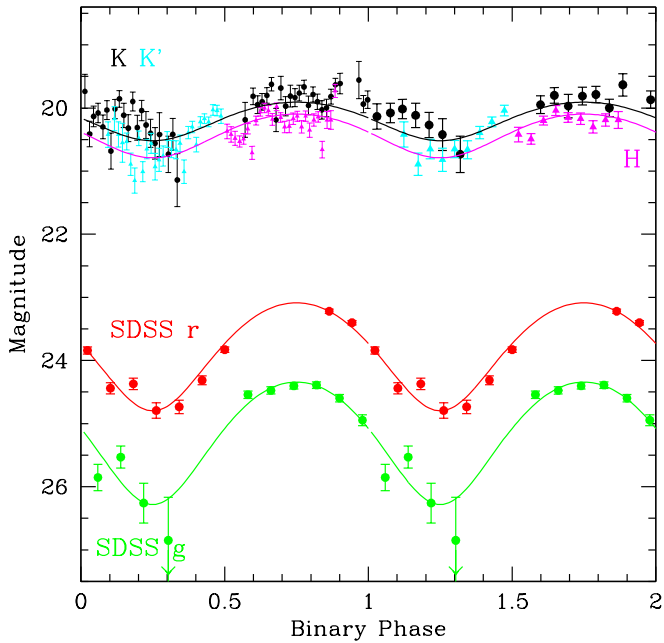}
\begin{center}
\caption{\label{LCs} 
GMOS-N/NIRI/NIRC2 Light curves.
The data are phased to the ephemeris of \citet{stoet14},
with $\phi=0$ at the pulsar ascending node; two periods are plotted for clarity.
For the second period, the infrared points are the medians of the magnitudes
during individual dither pointings. The model is an ICARUS direct heating fit to
the data.  Note the appreciable data fluctuations about the light curves, 
suggesting variability or finer substructure.
}
\end{center}
\vskip -0.4truecm
\end{figure}

\section{System Modeling and Conclusions}

	We fit these data with the ICARUS modeling code \citep{bet13}, optionally
including the extensions to compute the IntraBinary Shock (IBS) and its radiative heating \citep{rs16}
and the ducting of IBS particles to the companion surface by magnetic fields \citep{sr17}.
The extinction in this direction is modest, with the \citet{gsfet15} dust maps giving
$A_V\approx 0.22\pm0.06$ at 0.5\,kpc and $\approx 0.26\pm0.06$ by 1.0\,kpc. The basic fit considers
direct illumination by the pulsar. Lacking optical radial velocity information, we do not have enough
constraints for a meaningful fit to the neutron star mass, so it is fixed at
$M_{NS}=1.5M_\odot$. The primary fit parameters are the inclination $i$, the distance $d$, the Roche lobe fill factor
$f_{L1}$ (ratio of nose radius to the $L_1$ position, which determines the companion radius), 
$T_N$ the temperature of the unheated (Night) side of the companion and $L_D$ the direct
heating power.

	The basic fit is summarized in Table 1. A distance $d>1$\,kpc is preferred.
If $A_V$ is allowed to vary it tends toward zero, so we fix it at the lower limit for
the 1\,kpc distance.
The model requires a small inclination $i<30^\circ$ to match the shallow light curve. 
The fill factor gives a companion radius (volume equivalent Roche lobe radius) of 0.10$R_\odot$. 
The direct heating power $\sim 9 \times 10^{32} {\rm erg\,s^{-1}}$ is comfortably less than
the spindown power $5.6 \times 10^{33} {\rm erg\,s^{-1}}$. This fit is quite satisfactory except
for the relatively large $\chi^2/DoF \approx 2$, evidently the result of the substantial
scatter especially in the $K$ points.

\begin{deluxetable}{lrr}[t!!]
\tablecaption{\label{Fits} Binary Model Fits}
\tablehead{
\colhead{Parameter} & \colhead{Direct}& \colhead{IBS-B} 
}
\startdata
$i$ (deg)   &$24\pm 2$     & $40\pm6$  \cr 
$f_{L1}$    &$0.83\pm0.10$ & $0.97\pm0.08$    \cr 
$L_D/10^{32}\,{\rm erg\,s^{-1}}$ &$9.0\pm0.2$ &$8.0\pm0.4$ \cr
$L_P/10^{30}\,{\rm erg\,s^{-1}}^a$ &    &$1.1\pm0.4$        \cr
$T_N$ (K)   &$2420\pm220$  &$2650\pm340$      \cr 
$d$ (kpc)   &$1.06\pm0.05$ & $0.99\pm0.05$ \cr
$\chi^2\,(\chi^2/\nu)$&106 (2.04)  &84 (1.75) \cr
\\
$T_D$ (K)\tablenotemark{b} &3890 & 3880  \cr
$M_c ~(M_\odot)$ &$0.019$             &$0.013$  \cr
$R_c~(R_\odot)$\tablenotemark{c} &$0.104\pm0.013$  &$0.093\pm0.005$  \cr
$q$         &81.6       & 132.  \cr
$K$(km/s)\tablenotemark{d}   &250       & 399  \cr 
\enddata
\tablenotetext{a}{Magnetic axis at $\theta_B=22\pm 5^\circ$, $\phi_B=229\pm8^\circ$}
\tablenotetext{b}{Flux-averaged ``day''-side temperature}
\tablenotetext{c}{volume equivalent radius in $R_\odot$ for $f_{L1}$}
\tablenotetext{d}{Expected K-band radial velocity for $M_{NS}=1.5M_\odot$}
\end{deluxetable}

	In evaluating this fit, we should check consistency with other observables.
First the relatively large distance is consistent with the revised parallax. At this
distance the companion flux gives a radius $0.10 R_\odot$. Black widow companions
appear to be inflated by the heating and should hence be somewhat larger than 
the $R=0.0126 (M/M_\odot)^{-1/3}R_\odot$ radius of a cold (degenerate) H-poor remnant 
of a stellar core. With our fit inclination $i$ and the pulsar $x_2=a_2 {\rm sin} i = 0.00899 {\rm lt-s}$
one obtains a mass $M_C=0.019M_\odot$. This would have a minimum cold radius $0.047R_\odot$,
so the companion is appreciably inflated. Thus the observed rather low
$x_2$ and minimum companion mass is a result of the low inclination $i$;
the actual companion mass is quite typical of other BW pulsars.

	Note that the original parallax estimate of 203\,pc is impossible to accommodate 
in a direct heating model, since to match the observed companion flux and maximum temperature
the effective radius of the star would be 0.019$R_\odot$, less than half the
size of smallest plausible (cold degenerate) radius. This supports the general
conclusion of \citet{sr17} that direct heating models often indicate relatively 
large distances and luminosities. There we showed that if the heating is particle-
mediated and the particles precipitate to a magnetic cap, smaller luminosities and
distances are acceptable.

The absence of a source at this position in the 3FGL \citep{3FGL} or FL8Y surveys 
gives an upper limit on the gamma-ray flux (the dominant radiative output) of 
$f_\gamma < 5 \times 10^{-12} {\rm erg\,cm^{-2}s^{-1}}$. If the pulsar were an isotropic
emitter, then we would infer a gamma-ray heating power (at d=1.07\,kpc) 
$< 7 \times 10^{32} {\rm erg/s}$. This is slightly less than the required direct heating.
However, there is good evidence \citep[e.g.][]{wet09} that pulsar $\gamma$-rays are
preferentially beamed toward the spin equator. If, as expected, the pulsar
is spin-aligned with the binary orbit, at our inferred high inclination,
the Earth should detect a lower gamma-ray flux than that illuminating the companion.

	\citet{spet16} detect the pulsar system in a 15\,ks {\it XMM} exposure.
The limited X-ray counts do not allow a detailed spectral analysis, but with an assumed 
$N_H=3.3 \times 10^{20} \rm cm^{-2}$ (consistent with DM and $A_V$), they infer
a power law index $\Gamma=2.6$ and an unabsorbed flux
$f_{\rm 2-10\,keV} = 5.2 \times 10^{-15} {\rm erg\,cm^{-2}s^{-1}}$.
At the 203\,pc distance, this would have been an anomalously low X-ray luminosity,
but at our optically-fit distance the inferred $L_{\rm 2-10\,keV} = 7 \times 10^{29}
{\rm erg\,s^{-1}}$, is low but within the dispersion of the $L_X - {\dot E}$ relations
of \citet{poset02} and \citet{liet08}. As this $\Gamma$ is relatively
large for black widow non-thermal emission, the observed X-spectrum may be composite, with 
soft emission from a heated polar cap contributing at low energies. Deeper X-ray exposure will be
needed to make a detailed spectral measurement and to constrain possible
orbital modulation.

	The relatively large $\chi^2/DoF$ of the fit implies something must be present in
addition to the simple direct heating. IBS-accelerated particles
may be significant, and $\chi^2$ can be lowered at the cost of additional parameters.
Table 1 includes an example IBS fit, where in addition to direct heating we have 
a typical IBS front ($\beta=0.2$, ratio of companion wind momentum to pulsar wind
momentum) interacting with a companion magnetic field (dipole axis at
$\theta_B=29^\circ$ to the line of centers). Although the model does have
an improved $\chi^2/DoF$, the fit is not especially compelling as small
scale variations in the light curve are not matched. The general conclusions of the
direct heating fit are preserved, with a small inclination angle, acceptable
direct heating flux, $\sim\,$kpc distance and substantial filling factor
suggesting companion inflation. The particle heating is a small fraction of the
direct (gamma-ray) luminosity.

	The measured flux departures from the basic heating model look largely random.
Clearly we will require many orbits of high quality photometry to see what 
emission is variable and what is systematic. One intriguing feature is
the excess at $\phi=0.1-0.2$, with high points in $g$ and $r$ and a 
(variable?) excess in $K$. This is too narrow to be thermal emission
from a hot spot. However, if the IBS itself radiates sufficient optical
synchrotron flux, one may have narrow peaks from the emission beamed from the
relativistic plasma flowing on the IBS surface. Such emission would, however,
tends to be weak at the low $i$ indicated by the surface heating.

In the present fits, the high latitude view means that the Earth line-of-sight 
apparently passes above the equatorial ionized outflow from the companion wind, so no 
radio eclipses are seen. The small $i$ also implies a mass ratio $q \sim 100$
and a relatively large companion mass. This (despite the short orbital period) makes 
it more likely that binary is a classical black widow rather than the extreme mass 
ratio ($q\sim 200$) `Tidarren's like PSR J1311$-$3430 \citep{ret16}. 
With J0636's faint optical magnitude
and low temperature, a spectral check of the prime Tidarren signature (an H-free atmosphere) is
likely not feasible. However near-IR spectra might be able to provide a radial velocity, which
can help constrain the system mass and inclination; predicted values for the models
are given in the table. Higher S/N multicolor photometry, especially with multiple epochs
to isolate the quiescence light curve, can also help in understanding heating in this
very short-period system.

\bigskip
\bigskip

We thank N. Sanchez, who helped with the ICARUS code.
RWR was supported in part by NASA grant 80NSSC17K0024.

\end{document}